# AGILE Observations of the Gravitational Wave Event GW150914

M. Tavani[1,2,3], C. Pittori[4,5], F. Verrecchia[4,5], A. Bulgarelli[6], A. Giuliani[7], I. Donnarumma[1], A. Argan[1], A. Trois[8], F. Lucarelli[4,5], M. Marisaldi[6], E. Del Monte[1], Y. Evangelista[1], V. Fioretti[6], A. Zoli[6], G. Piano[1], P. Munar-Adrover[1], L.A. Antonelli[4,5], G. Barbiellini[9], P. Caraveo[7], P.W. Cattaneo[10], E. Costa[1], M. Feroci[1], A. Ferrari[11], F. Longo[9], S. Mereghetti[7], G. Minervini[12], A. Morselli[13], L. Pacciani[1], A. Pellizzoni[8], P. Picozza[13], M. Pilia[8], A. Rappoldi[10], S. Sabatini[1], S. Vercellone[14], V. Vittorini[1], P. Giommi[4], S. Colafrancesco[15], M. Cardillo[16].

[1]INAF-IAPS, via del Fosso del Cavaliere 100, I-00133 Roma, Italy

[2]Dip. di Fisica, Univ.di Roma "Tor Vergata", via della Ricerca Scientifica 1, I-00133 Roma, Italy

[3]Gran Sasso Science Institute, viale Francesco Crispi 7, I-67100 L'Aquila, Italy

[4]ASI Science Data Center (ASDC), Via del Politecnico, I-00133 Roma, Italy

[5]INAF-OAR, via Frascati 33, I-00078 Monte Porzio Catone (Roma), Italy

[6]INAF-IASF-Bologna, Via Gobetti 101, I-40129 Bologna, Italy

[7]INAF-IASF Milano, via E.Bassini 15, I-20133 Milano, Italy

[8]INAF, Osservatorio Astronomico di Cagliari, Poggio dei Pini, strada 54, I-09012 Capoterra, Italy

[9]Dip. di Fisica, Universita' di Trieste and INFN, Via Valerio 2, I-34127 Trieste, Italy

[10]INFN-Pavia, Via Bassi 6, I-27100 Pavia, Italy

[11]CIFS, c/o Physics Department, University of Turin, via P. Giuria 1, I-10125, Torino, Italy

[12]Dip. di Matematica, Univ.di Roma "Tor Vergata", via della Ricerca Scientifica 1, I-00133 Roma, Italy

[13]INFN Roma Tor Vergata, via della Ricerca Scientifica 1, I-00133 Roma, Italy

[14]INAF/IASF-Palermo, Via U.La Malfa 153, I-90146 Palermo, Italy

[15] University of Witwatersrand, Johannesburg, South Africa

[16]INAF Osservatorio Astronomico di Arcetri, Largo Enrico Fermi, 5, I-50125 Firenze, Italy

## ABSTRACT

We report the results of an extensive search in the AGILE data for a gamma-ray counterpart of the LIGO gravitational wave event GW150914. Currently in spinning mode, AGILE has the potential of covering with its gamma-ray instrument 80% of the sky more than 100 times a day. It turns out that AGILE came within a minute from the event time of observing the accessible GW150914 localization region. Interestingly, the gamma-ray detector exposed $\sim 65$ % of this region during the 100 s time intervals centered at -100 s and +300 s from the event time. We determine a $2\sigma$ flux upper limit in the band 50 MeV - 10 GeV, $UL = 1.9 \times 10^{-8} \, \mathrm{erg \, cm^{-2} \, s^{-1}}$ obtained $\sim 300$ s after the event. The timing of this measurement is the fastest ever obtained for GW150914, and significantly constrains the electromagnetic emission of a possible high-energy counterpart. We also carried out a search for a gamma-ray precursor and delayed emission over timescales ranging from minutes to days: in particular, we obtained an optimal exposure during the interval $-150/-30$ s. In all these observations, we do not detect a significant signal associated with GW150914. We do not reveal the weak transient source reported by *Fermi*-GBM 0.4 s after the event time. However, even though a gamma-ray counterpart of the GW150914 event was not detected, the prospects for future AGILE observations of gravitational wave sources are decidedly promising.

*Subject headings:* gravitational waves, gamma rays: general.

## 1. Introduction

The recent discovery of gravitational waves (hereafter, GW) by the LIGO experiment impulsively emitted from the source GW150914 started a new era in astronomy (Abbott et al., 2016a,b,c,d; hereafter A16a,b,c,d). The detection occurred at the beginning of an acquisition run of the LIGO experiment in an enhanced configuration (A16a). The LIGO-VIRGO detectors are expected to operate soon at even improved sensitivity, and the potential for a large number of detections of gravitational phenomena will shape future ground-based and space research. The characteristics of gravitational phenomena emitting detectable gravitational waves are those of the most extreme and energetic events of our universe. The energy radiated under the form of gravitational waves as inferred from GW150914 is about 3 $M_\odot$, a huge value (A16a). This energy was emitted during a few hundreds of milliseconds. It is clear that this type of gravitational phenomena opens the way to study with unprecedented synergy the interplay between gravitation, the astrophysical context, and the quantum properties of fields and matter. In general, GW-emitting final stages of coalescences involving compact stars (neutron stars (NSs) and black holes (BHs)) are the most likely candidates for LIGO-VIRGO events. X-ray and gamma-ray counterparts are expected from different types of coalescing compact star systems. The quest for electromagnetic counterparts of extreme gravitational events is now open.

The characteristics of the binary system associated with GW150914 are somewhat surprising given the current observations and understanding of the evolutionary processes leading to the formation of black holes. The event results from the coalescence of two black holes of relatively large masses (near 30 $M_{\odot}$) in an unknown stellar environment (A16d). Where and how such events can be produced is an open and interesting question (A16d): the physical conditions of the coalescing compact objects are far from being understood. Even though purely gravity-systems such as BH-BH binaries are not anticipated to emit detectable electromagnetic (e.m.) radiation, nevertheless such a radiation can be emitted before, during and after coalescence depending on the physical conditions of the system. It is then of great interest to explore this possibility and search for e.m. counterparts of GW events.

The AGILE satellite, today at its ninth year of operations in orbit, is observing the gamma-ray sky with excellent monitoring capabilities in the gamma-ray range 30 MeV - 30 GeV. The satellite is currently in spinning mode covering a large fraction of the sky with a gamma-ray sensitivity to transient emission that can reach flux levels near $F = (1-2) \times 10^{-8}\,\mathrm{erg\,cm^{-2}\,s^{-1}}$ for ~100 s integrations. This timescale is typical of passes of the imaging gamma-ray instrument (Field of View (FoV) of 2.5 sr) over exposed sky regions in spinning mode. Each accessible sky region is exposed more than 100 times a day[1] with 100 s integrations each. In addition, non-imaging detectors such as the AGILE Mini-Calorimeter and anticoincidence system (routinely employed for the detection of GRBs and TGFs) can be used. AGILE has therefore unique capabilities for the search of gamma-ray counterparts of GW sources.

In this *Letter* we present the results of an extensive search in AGILE data for a high-energy counterpart of GW150914 that we performed retroactively once the event was made public[2]. The paper is organized as follows. In Sect. 2 we state the case of AGILE as an ideal space instrument for the search of gamma-ray counterparts of GW events. In Sect. 3 we present the results of AGILE observations and extended analysis regarding the GW150914 event. In Sects. 4 and 5 we briefly discuss our results and future perspectives.

---

[1]Depending on Earth occultations and SAA passages.

[2]The AGILE Team was not part of the multifrequency follow-up collaboration with the LIGO-VIRGO team at the time of the GW150914 detection and subsequent investigations; we learned of the GW150914 event on February 11, 2016.

## 2. AGILE's Capability for the Search of Gamma-Ray Counterparts of GW Sources

The AGILE satellite, launched on April 23rd, 2007, is orbiting the Earth in a near equatorial orbit of current altitude $\sim 500$ km. The instrument consists of an imaging gamma-ray Silicon Tracker (sensitive in the energy range 30 MeV - 30 GeV), Super-AGILE (currently working in ratemeter mode in the energy range 20-60 keV due to current temporary telemetry limitations), a Mini-Calorimeter (MCAL, working in the range 0.4 - 100 MeV) and an anticoincidence (AC) system (for a summary of the AGILE mission features, see Tavani et al. 2009). The combination of Tracker, MCAL and AC working as a gamma-ray imager constitutes the AGILE-GRID. The instrument is capable of detecting gamma-ray transients and GRB-like phenomena for timescales ranging from sub-milliseconds to tens-hundreds of seconds. In addition to the hundreds of GRBs detected by MCAL and Super-AGILE, several prominent GRBs were detected by the gamma-ray imager since the beginning of operations (GRB080514B, Giuliani et al 2008; GRB090401B, Moretti et al. 2009; GRB090510, Giuliani et al. 2010; GRB100724B, Del Monte et al. 2011; GRB130327B, Longo et al. 2013; GRB130427A, Verrecchia et al. 2013; GRB131108A, Giuliani et al. 2013 and Giuliani et al. 2015). Furthermore, AGILE so far detected about 1,000 Terrestrial Gamma-Ray Flashes (TGFs) with durations ranging from hundreds to thousands of microseconds (Marisaldi et al. 2014, Tavani et al. 2011). A special sub-millisecond search for transient events detected by MCAL is operational on board (Tavani et al., 2009).

The characteristics that make AGILE in spinning mode an important instrument for follow-up observations of large GW source localization regions are: (1) a very large FoV of the GRID (2.5 sr); (2) an accessible region of 80% of the whole sky that can be exposed every 7 minutes (see Fig. 1); (3) 100-150 useful passes every day for any region in the accessible sky[3]; (4) a gamma-ray exposure of $\sim 2$ minutes of any field in the accessible sky every 7 minutes with a sensitivity reaching $\sim 10^{-8}\,\mathrm{erg\,cm^{-2}\,s^{-1}}$ above 30 MeV for typical single-pass of unocculted sky regions; (5) sub-millisecond trigger for very fast events detectable by MCAL in the range 0.4-100 MeV; (6) hard X-ray $(20-60\ \mathrm{keV})$ triggers of GRB-like events with a localization accuracy of 2-3 arcmin in the Super-AGILE FoV ($\sim 1$ sr) when operating in imaging mode.

---

[3]The total number of 7 minute rotations is $\sim 200$/day; the useful number for gamma-ray exposure is affected by Earth occultations and SAA passages.

Satellite data are transmitted to the ground currently on average every two consecutive passes over the Malindi ground station in Kenya. Scientific data are then processed by fast processing, typically producing alert for transient gamma-ray source and/or GRB-like events within 1.5-3 hrs from satellite on-board acquisition (Bulgarelli et al. 2014, Pittori et al 2013).

## 3. AGILE Observations of GW150914

The GW150914 event occurred at time $T_0 = $ 09:50:45 UTC on Sept. 14, 2015 (A16a). At that time AGILE was scanning the sky in spinning mode with the Earth only partially occulting the GRID FoV. Fig. 2 shows the gamma-ray exposure above 50 MeV for the whole satellite 7 minute revolution that includes the GW150914 event time. As anticipated, the Earth only marginally covers the GW150914 localization region of the most accurate GW150914 localization map (LALInference, Abbott et al. 2016e, Veitch et al. 2015): most of the localization region is not occulted, and therefore available for AGILE exposure. We performed a search for: (1) the prompt event involving the GRID, MCAL, AC and Super-AGILE ratemeters (both inside and outside the GRID FoV); (2) delayed emission on multiple timescales involving the GRID; (3) precursor emission involving the GRID.

### 3.1. The prompt event

Figure 3 shows the gamma-ray exposure of a typical point inside the GW150914 localization region for the two satellite rotations of interest closer to the prompt event. AGILE had GRID exposure of a substantial fraction (65%) of the GW150914 localization region a few tens of seconds before $T_0$, but not at the event time (see also Fig. 4). Had the instrument obtained an exposure of the field equal to that of a few tens of seconds earlier, AGILE could have obtained a gamma-ray sensitivity near $10^{-6}\,\mathrm{erg\,cm^{-2}\,s^{-1}}$ for a few second integration in the range 50 MeV - 10 GeV. A large fraction of the GW localization region was not occulted by the Earth and a strong X/gamma-ray signal, if any, could have been detected by the AGILE non-imaging detectors.

The MCAL did not trigger events within an interval covering -100 / +100 seconds from $T_0$. Within -200 / +200 seconds we found four triggers, all of them on the sub-millisecond or $\sim 1$ ms timescales. Considering the events' characteristics, three apparent transients out of four can be attributed to instrumental noise, while one is a candidate TGF. MCAL did not detect the transient reported by the *Fermi* GBM team (Connaughton et al., 2016) at $T_0 + 0.4s$. For the reported spectral shape (best-fit single power-law with index $-1.4^{+0.18}_{-0.24}$) we estimate a minimum detectable MCAL fluence at the standard trigger threshold (five standard deviations above background) of $\sim 1.0 \times 10^{-6}$ erg cm$^{-2}$ in the $400 - 10000$ keV energy range, for a 1-s exposure at 70° off axis. This is only 13% larger than the fluence reported for the *Fermi*-GBM event $(2.4^{+1.7}_{-1.0} \times 10^{-7}$ erg cm$^{-2}$ in the $10 - 1000$ keV energy range), when the difference in the energy range is accounted for.

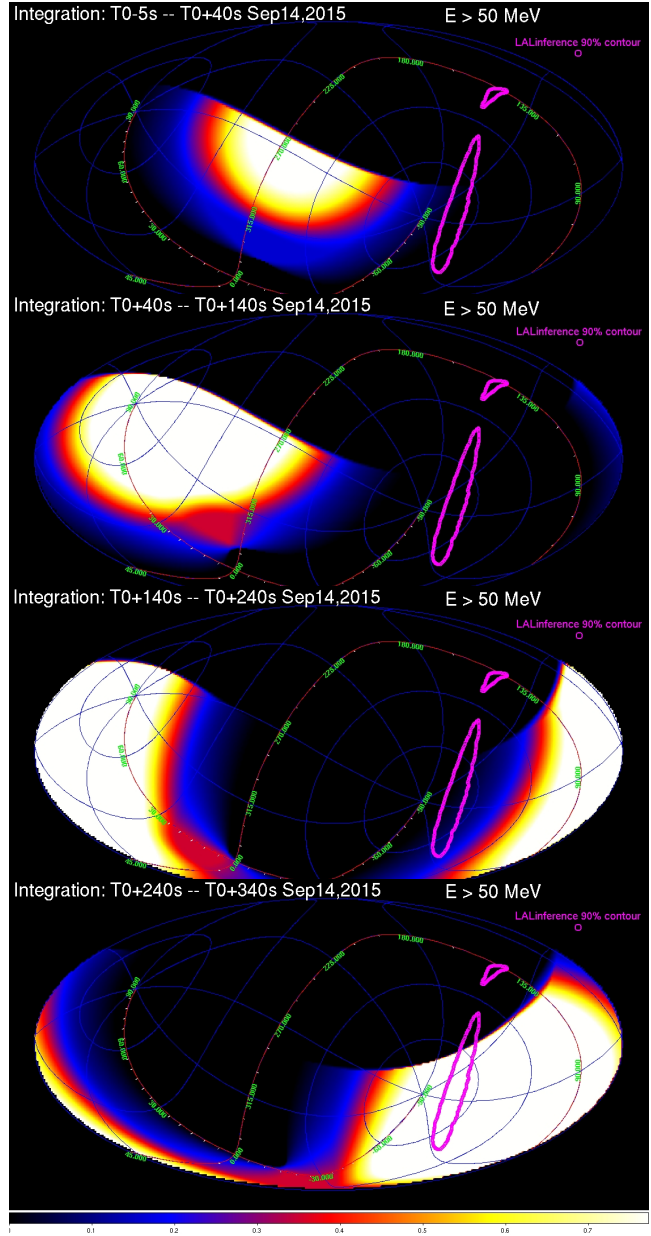

Fig. 1.— Hammer-Aitoff projection in Galactic coordinates of the AGILE satellite rotation and a typical sequence of $E > 50$ MeV gamma-ray exposure maps (in $\mathrm{cm^2\,s\,sr}$, using a $0.5°$ pixel size) as the satellite rotates in spinning mode scanning 80% of the sky in about 7 minutes. The AGILE-GRID FoV radius is assumed to be $70°$. In this case, the Earth is occulting the FoV in the Northern Galactic hemisphere. The exclusion region for the albedo photons is $80°$ from the Earth center. This sequence applies to the satellite rotation that includes the prompt event time of GW150914 (localization region marked by the purple contour, LALinference 90% contour level, A16e).

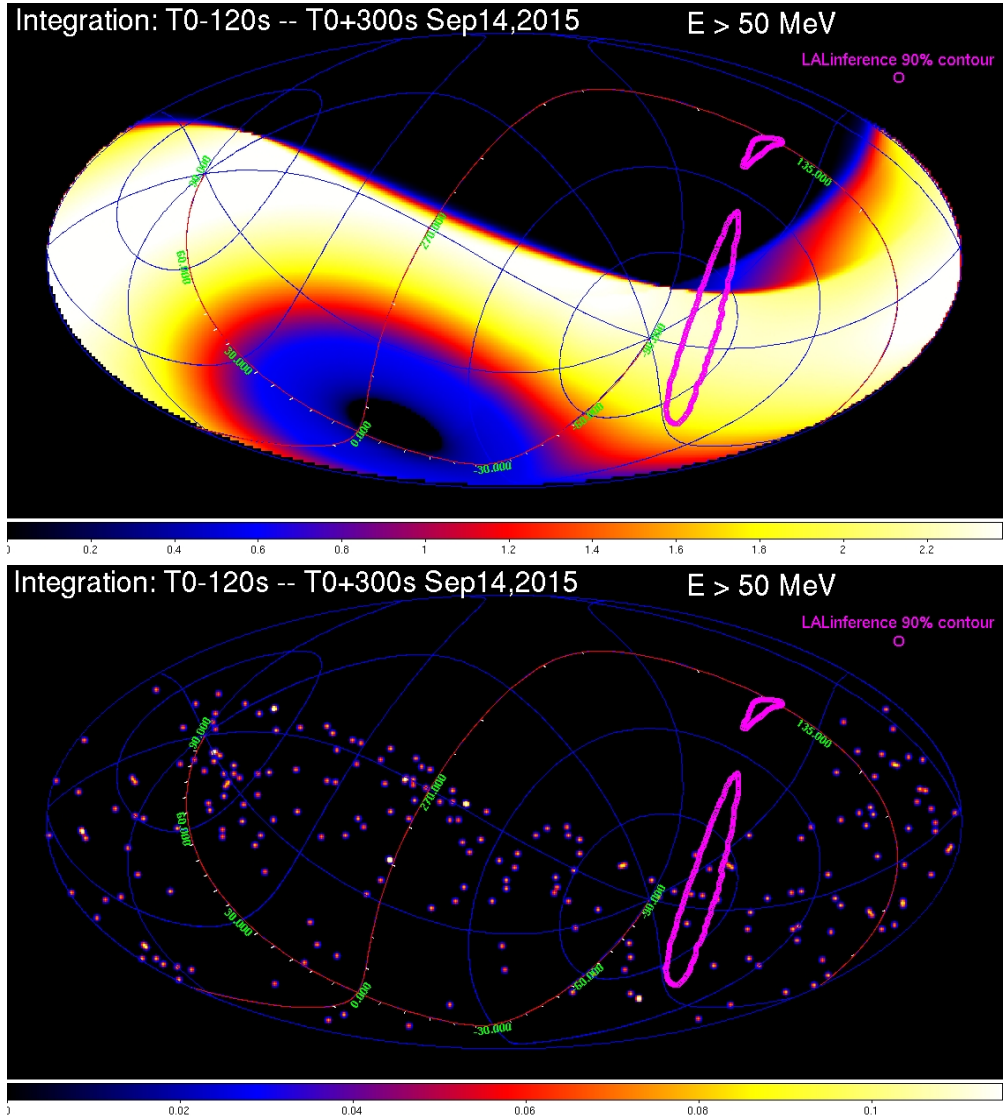

Fig. 2.— Hammer-Aitoff projection in Galactic coordinates. *Top panel:* AGILE cumulative gamma-ray exposure above 50 MeV (in cm² s sr, using a 0.5° pixel size) during the satellite 7 minute rotation that includes the GW150914 event time. *Bottom panel:* Gamma-ray photons above 50 MeV detected during the satellite 7 minute rotation. The GW150914 localization region is marked by the area inside the purple line (LALinference 90% contour level, A16e).

A search in the Super-AGILE ratemeters data does not produce a significant detection with a $2\sigma$ fluence upper limit of $2.4 \times 10^{-8}$ erg cm$^{-2}$. We remind here that the interesting short GRB090510 (Giuliani et al. 2010, see discussion below) is detected by the SA ratememeters with a significance level of $\sim 8.5\,\sigma$ although it occurred outside the 1 sr SA field of view. Also a search of a signal in AC data does not produce a significant detected flux.

AGILE was optimally positioned in the GW150914 localization region at interesting time intervals preceding and following the prompt event. The most interesting time intervals are during the time intervals $\Delta T_{-1} = -95 \pm 50$ s, and $\Delta T_{+1} = +333 \pm 50$ s, taking $T_0 = 09:50:45$ UTC on Sept. 14, 2015 as a time-zero reference. In the following, we focus on the analysis of the gamma-ray data searching for precursor and delayed gamma-ray emission. Table 1 shows the AGILE-GRID passes over the GW150914 localization region and the results of our analysis in search of transient gamma-ray emission in that region.

### 3.2. A search for Delayed Emission

The AGILE-GRID exposed a good fraction (75%) of the GW150914 localization region within 250 s from the prompt event. As shown in Figs. 3 and 5, important information can be obtained during the first useful pass, $\Delta T_{+1}$. Considering the local photon background and exposure, a search in the localization region for a transient gamma-ray source produces the $2\sigma$ UL for emission[4] in the range 50 MeV - 10 GeV: $UL = 1.9 \times 10^{-8}$ erg cm$^{-2}$ s$^{-1}$. This upper limit is significant in the context of a possible gamma-ray signal from a cosmic event associated with compact object coalescence. Figure 6 shows the upper limit obtained by AGILE in the context of the gamma-ray lightcurve expected from the short GRB090510 repositioned at redshift $z = 0.09$ of GW150914. This short GRB, which shows several features expected from compact object coalescences possibly emitting GWs (e.g., Berger 2014), was detected by AGILE and *Fermi*-LAT with a significant delayed emission above 30 MeV lasting up to 100 s with a hard spectral component (Giuliani et al. 2010, Abdo et al. 2009, and

---

[4]For a hard spectrum similar to the short GRB delayed emission of GRB090510 discussed below.

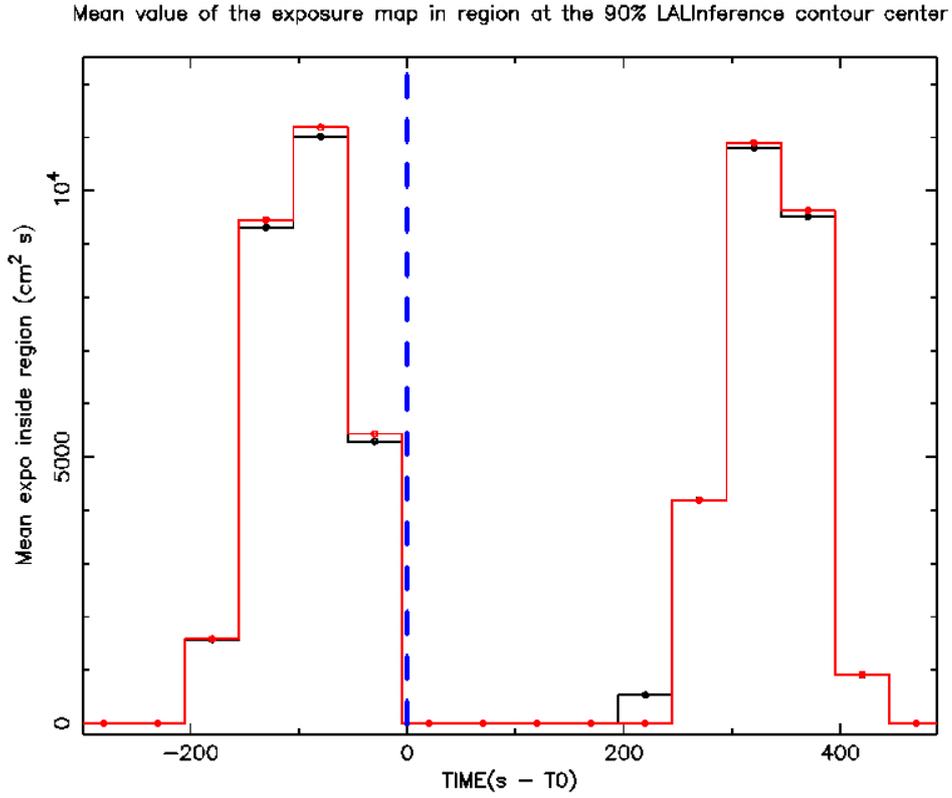

Fig. 3.— AGILE gamma-ray exposure above 50 MeV as a function of time of a region positioned at the center of the GW150914 localization region for the two passes, one preceding and one following $T_0$ of GW150914. The GW150914 event time is marked by a dotted blue line. The black curve is obtained for a $7° \times 25°$ field centered at Galactic coordinates $(l, b) = (282.8, -24.6)$. The red curve is obtained for a $10°$ radius circle centered at Galactic coordinates $(l, b) = (283.5, -25.8)$.

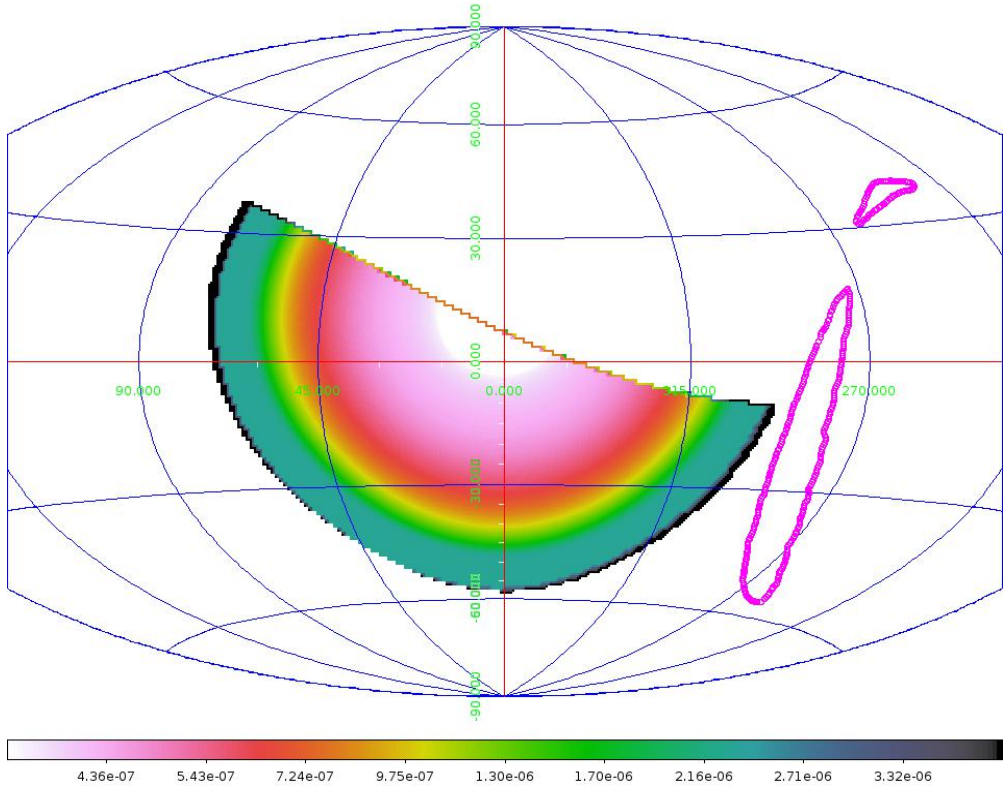

Fig. 4.— Hammer-Aitoff projection in Galactic coordinates of the AGILE gamma-ray sensitivity above 50 MeV (in $\mathrm{erg\,cm^{-2}\,s^{-1}}$) during the 4 s interval that includes the GW150914 event time. The AGILE-GRID FoV is taken to be 70°. AGILE just missed covering the GW150914 localization region during the prompt event with useful sensitivity. The GW150914 localization region is marked by the purple contour (LALinference 90% contour level, A16e).

**Table 1: Analysis of individual passes over the GW150914 localization region**

| Interval number | Central time bin (**) | Duration (s) | $2\sigma$ UL (*) ($10^{-8}\,\mathrm{erg\,cm^{-2}\,s^{-1}}$) | Comments |
|---|---|---|---|---|
| -13 | -5203 | 100 | 2.7 | 88% of error box not-occulted |
| -12 | -4779 | 100 | – | affected by SAA |
| -11 | -4355 | 100 | – | affected by SAA |
| -10 | -3931 | 100 | – | affected by SAA |
| -9 | -3507 | 100 | – | affected by SAA |
| -8 | -3083 | 100 | 2.3 | 93% of error box not-occulted |
| -7 | -2663 | 100 | 4.5 | 78% of error box not-occulted |
| -6 | -2235 | 100 | 1.5 | 68% of error box not-occulted |
| -5 | -1807 | 100 | 1.5 | 65% of error box not-occulted |
| -4 | -1379 | 100 | 1.5 | 20% of error box not-occulted |
| -3 | -951 | 100 | 1.0 | 48% of error box not-occulted |
| -2 | -523 | 100 | 1.0 | 56% of error box not-occulted |
| -1 | -95 | 100 | 1.5 | 65% of error box not-occulted |
| +1 | +333 | 100 | 1.9 | 75% of error box not-occulted |

(*) Flux upper limit obtained for emission in the range 50 MeV - 10 GeV and for a spectrum similar to the delayed gamma-ray emission of the short GRB090510.

(**) Time calculated from $T_0 = 09:50:45$ UTC on Sept. 14, 2015, the event time of GW150914.

**Table 2: Long-integration time analysis of the GW150914 localization region**

| Interval name | Duration | $2\sigma$ UL (*) ($10^{-9}\,\mathrm{erg\,cm^{-2}\,s^{-1}}$) | Comments |
|---|---|---|---|
| -3d | 3 days | 0.3 | |
| -2d | 2 days | 0.5 | |
| -1d | 1 day | 0.7 | |
| -12h | 12 hours | 0.8 | |
| -6h | 6 hours | 2.5 | |
| -3h | 3 hours | 3.5 | |
| +3h | 3 hours | – | telemetry interruption (**) |
| +6h | 6 hours | 3.5 | with telemetry interruption (**) |
| +12h | 12 hours | 1.8 | with telemetry interruption (**) |
| +1d | 1 day | 1.1 | with telemetry interruption (**) |
| +2d | 2 days | 0.9 | with telemetry interruption (**) |
| +3d | 3 days | 0.7 | with telemetry interruption (**) |
| +5d | 5 days | 0.4 | with telemetry interruption (**) |

(*) Gamma-ray upper limits (ULs) for photons in the range 100 MeV - 10 GeV and for a $E^{-2}$ spectrum obtained in the best exposed regions of the GW150914 localization region. We established a range of ULs within a factor of 2 inside the exposed localization region. The ULs were obtained taking into account the effective exposure that has an overall efficiency near 70% due to Earth occultation and SAA passages.

(**) Telemetry interruption from UT 10:00 to UT 13:00 of Sept. 14, 2015.

Ackermann etal. 2010). Thus AGILE data uniquely determine a gamma-ray upper limit within 250-350 s from the GW150914 event.

Due to a temporary telemetry interruption caused by ground station operations in Malindi (Kenya), satellite data for two consecutive orbits following the interval $\Delta T_{+1}$ are not available. The next useful pass over the GW150914 localization region is about 3 hours after $\Delta T_{+1}$, at $T = 10,800$ s after $T_0$.

We carried out a long-timescale search for transient gamma-ray emission during the hours immediately following the prompt event. No significant gamma-ray emission in the GW150914 localization region was detected during individual passes 3-4 hours after the GW150914 event.

We also performed a search on longer timescales up to several days after the event. Table 2 summarizes our results and upper limits for these long integrations.

### 3.3. A search for a gamma-ray precursor

An interesting possibility arises from the situation of two approaching massive BHs in a shrinking orbit. Even though current theoretical scenarios for BH-BH coalescences do not envision gamma-ray emission preceding or following the final event (e.g., Baumgarte & Shapiro 2011), a residual gaseous and/or plasma environment can in principle induce e.m. radiation during the approaching phase by non-thermal processes. It is in any case relevant determining if gamma rays were emitted before the coalescence itself.

As reported in Tables 1 and 2, we carried out a search for a gamma-ray precursor over a large dynamic time range. The most significant observations are the available AGILE-GRID passes preceding the prompt event. Table 1 and Fig. 5 show the sequence of passes. One of the most interesting passes is $\Delta T_{-1}$ covering the time interval $-95 \pm 50$ s from $T_0$. The gamma-ray $2\sigma$ UL in this case is comparable with what obtained for the $\Delta T_{+1}$ interval, that is $UL = 1.5 \times 10^{-8}\,\mathrm{erg\,cm^{-2}\,s^{-1}}$. As Fig. 5 shows, for the successive passes (retrograde in time) -2 and -3, the Earth progressively enters in the exposed GW150914 localization region. Passes -4 and -5 have a marginal exposure with a relatively high background because of the Earth albedo gamma-ray photons. The relevant region is then better exposed for intervals -6, -7, and -8. Interval -8 is one of the best cases, with more than 90% of the GW150914 localization region region well sampled. Unfortunately, the successive intervals -9, -10, -11, -12 are affected by a particularly deep passage in the SAA[5] and no useful data are available for these intervals. Pass no. -13 is very similar to the pass +1, and occurs after one orbit ($\sim 95$ minutes).

We also extended retrogradely in time our search up to 3 days before $T_0$. No significant gamma-ray emission was detected during any time interval reported in Tables 1 and 2.

### 4. Discussion

AGILE observed the field containing GW150914 with very good coverage and significant gamma-ray exposure within tens-hundreds of minutes before and after the event. The AGILE-GRID missed the coverage of the prompt event in its field of view, but could determine important limits immediately before and after the event. In light of the broad campaign of follow-up observations of the GW150914 localization region ranging from radio to gamma-rays reported in Abbott et al. 2016e, the AGILE observations are significant in providing the fastest response to the event with optimal gamma-ray sensitivity.

---

[5]Usually, passages into the SAA affect 1-2 rotations. In the period near $T_0$ of GW150914 , the SAA passage was "deep" and affected 4 satellite rotations.

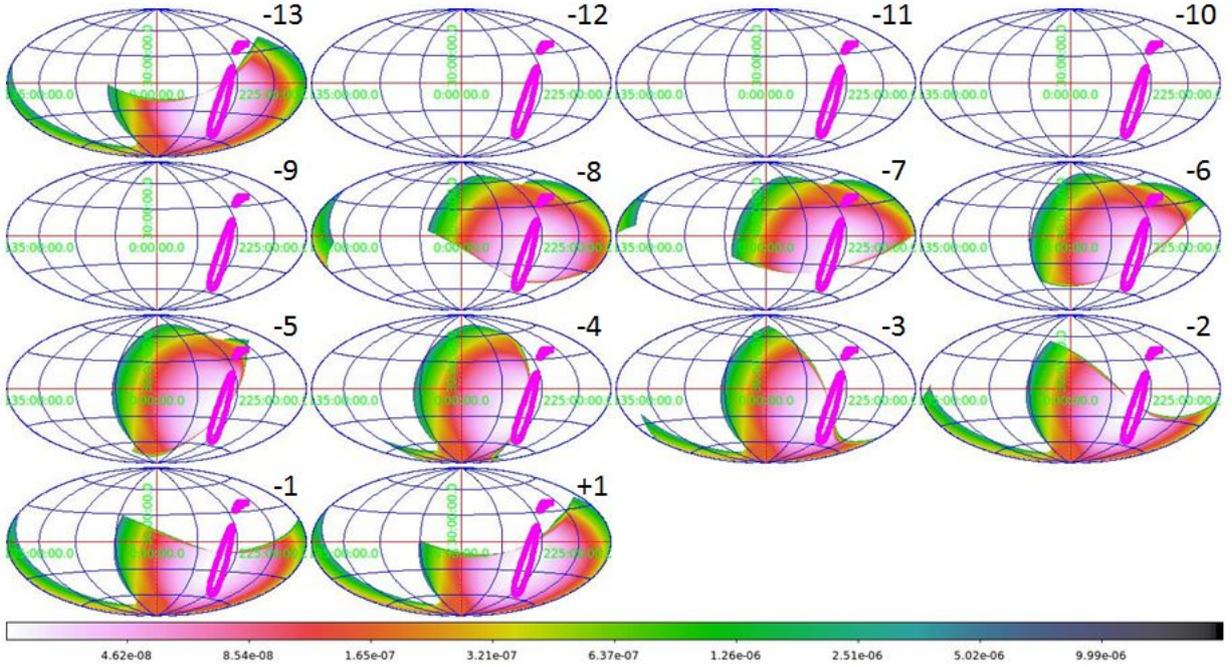

Fig. 5.— Sequence of ($E > 50$ MeV) maps in Galactic coordinates showing the AGILE-GRID passes with the best sensitivity over the GW150914 localization region obtained during the period (- 5,303 s, + 433 s) with respect to $T_0$. The color maps show the gamma-ray flux $2\sigma$ upper limits in the range 50 MeV - 10 GeV with the most stringent values being $UL = (1 - 2) \times 10^{-8}$ erg cm$^{-2}$ s$^{-1}$. The sequence shows 14 maps for all the 1-orbit passes of Table 1, corresponding to the 100 s interval numbers (from top left to bottom right): $-13, -12, -11, -10, -9, -8, -7, -6, -5, -4, -3, -2, -1, +1$. Note that the passes $-12, -11, -10, -9$ are blank (i.e., do not have appreciable exposure and therefore no reported ULs) because they are strongly affected by the SAA. The GW150914 localization region is marked by the purple contour (LALinference 90% contour level, A16e).

The most important result of the AGILE observations is the gamma-ray UL close to $10^{-8}\,\mathrm{erg\,cm^{-2}\,s^{-1}}$ in the range 50 MeV - 10 GeV obtained in the interval 250-350 s after $T_0$. It is interesting to compare this UL with other observations obtained by imaging X-ray/gamma-ray space instruments with FoVs larger that 1 sr. The BAT instrument on board of the *Swift* satellite could not observe the GW150914 field because it occurred outside its field of view (Evans et al. 2016). The same applies to *Fermi*-LAT that could not cover the GW150914 localization region at the moment of the prompt event (*Fermi*-LAT Collaboration, 2016). Compared with the *Fermi*-LAT gamma-ray UL that was obtained more than 70 minutes after the event (*Fermi*-LAT Collab., 2016), the AGILE-GRID observation at $\Delta T_{+1}$ provides a more stringent constraint to any delayed emission above 50 MeV shortly after the event.

We note that "delayed" gamma-ray emission is sometimes detected from both long and short GRBs (e.g., Giuliani et al. 2010, Ackermann et al. 2010, De Pasquale et al. 2010). In particular, the case of the gamma-ray bright short GRB 090510 is relevant to our purpose because of its characteristics and possible association with a compact star coalescence involving GW emission. The event, localized through its afterglow at $z = 0.9$, showed a first quasi-thermal interval lasting about 300 ms with a spectrum peaking at a few MeV, and a second "gamma-ray afterglow" phase lasting hundreds of seconds detected above 100 MeV with a highly non-thermal spectrum (Giuliani et al. 2010, Ackermann et al. 2010). The possibility then of detected delayed gamma-rays from GRBs is of great relevance for detectors such as the AGILE-GRID. As shown in Fig. 6 the AGILE UL obtained 300 seconds after the GW event is just below the gamma-ray emission expected from a short GRB090510-like event re-located at the distance of $z = 0.1$. Our data are close to excluding a delayed gamma-ray afterglow of the type of GRB090510. We notice that a short GRB of this kind positioned at 400 Mpc would have been detectable by the AGILE non-imaging detectors (AC, MCAL and Super-AGILE ratemeters) as well as by *Fermi*-GBM and INTEGRAL/SPI-ACS if the event had the same hard X-ray spectrum as GRB090510. We checked that the current sensitivity of the AGILE AC system is about a factor of 5-10 better than the flux detected in the case of the short GRB 090510, implying that a signal 10 times weaker than that associated with GRB090510 would have been detected.

Also of interest is the limit obtained 50-100 s before the coalescence. For a total mass near $60\,M_\odot$, we are sampling the radiative environment when the orbital distance $a$ is $\sim (0.1-1)R_\odot$. We are excluding precursor gamma-ray activity at the distance comparable with the solar size during the compact object approach. If dynamically formed in a dense stellar environment, the BH-BH binary associated to GW150914 might carry a gaseous remnant with it as a product of the 3-body encounter that formed the binary. This gaseous component subject to the gravitational influence of the approaching massive BHs can settle into a short-lived disk that might produce e.m. radiation by thermal and non-thermal processes. The observed flux UL translates into an upper limit to the (isotropically) radiated gamma-ray luminosity, $L_\gamma < 3 \times 10^{47}\,\mathrm{erg\,s^{-1}}$, assuming the GW150914 luminosity distance of $\sim 400$ Mpc. For a duration of 100 s, this corresponds to a limit to the radiated gamma-ray energy, $E_\gamma \sim 1.7 \times 10^{-5}\,M_\odot$.

AGILE does not detect the very weak event reported by *Fermi*-GBM about 0.4 s after $T_0$ (Connaughton etal., 2016). The MCAL did not register any event above the trigger threshold in the energy range 0.4-100 MeV, and the AC and Super-AGILE detectors did not register enhancements of their countrates during the GW150914 prompt event (in the ranges 80-200 keV and 20-60 keV, respectively). Considering the faintness and spectrum of the *Fermi*-GBM event[6] this lack of detection is not surprising.

## 5. Future Observations

In the near future, we expect a relatively large number of aLIGO and aLIGO-VIRGO detections of GW sources. As the GW sensitivity and event positioning is improving, the capability to alert the community within timescales progressively short is to be envisioned. With aLIGO-aVIRGO being capable of detecting NS-NS and NS-BH coalescing systems, the likelihood of detecting electromagnetic emission from these events will increase substantially.

We showed in this paper that AGILE can effectively observe relatively large sky regions associated with GW events with high efficiency. As demonstrated in the case of GW150914, AGILE might obtain significant results if the event occurs in the accessible sky region which depends on the sky coverage per 7 minute rotation ($\sim 0.8$) and Earth occultation ($\sim 0.6$), with a probability $\sim 0.8 \times 0.6 \simeq 0.5$. In this case, AGILE can obtain imaging gamma-ray data within a timescale ranging from seconds to 300-400 s. This is exactly the case that occurred for GW150914. The GRID FoV just missed the prompt event, and had exposure for most of the localization region only $\sim$300 s after $T_0$. As shown in Sect. 2, this is the worst case that could have happened given the conditions. Nevertheless, the upper limit obtained for this "late" exposure provides the fastest data in constraining the GW150914 e.m. emission.

---

[6]This latter event was also not confirmed by the INTEGRAL SPI-ACS observation covering the GW150914 prompt event (Savchenko et al. 2016). The SPI-ACS upper limit is important since it is obtained by another satellite not occulted by the Earth at the time of $T_0$. The analysis of the SPI-ACS instrument is reported in the interval from $-30$ s to $+30$ s from $T_0$.

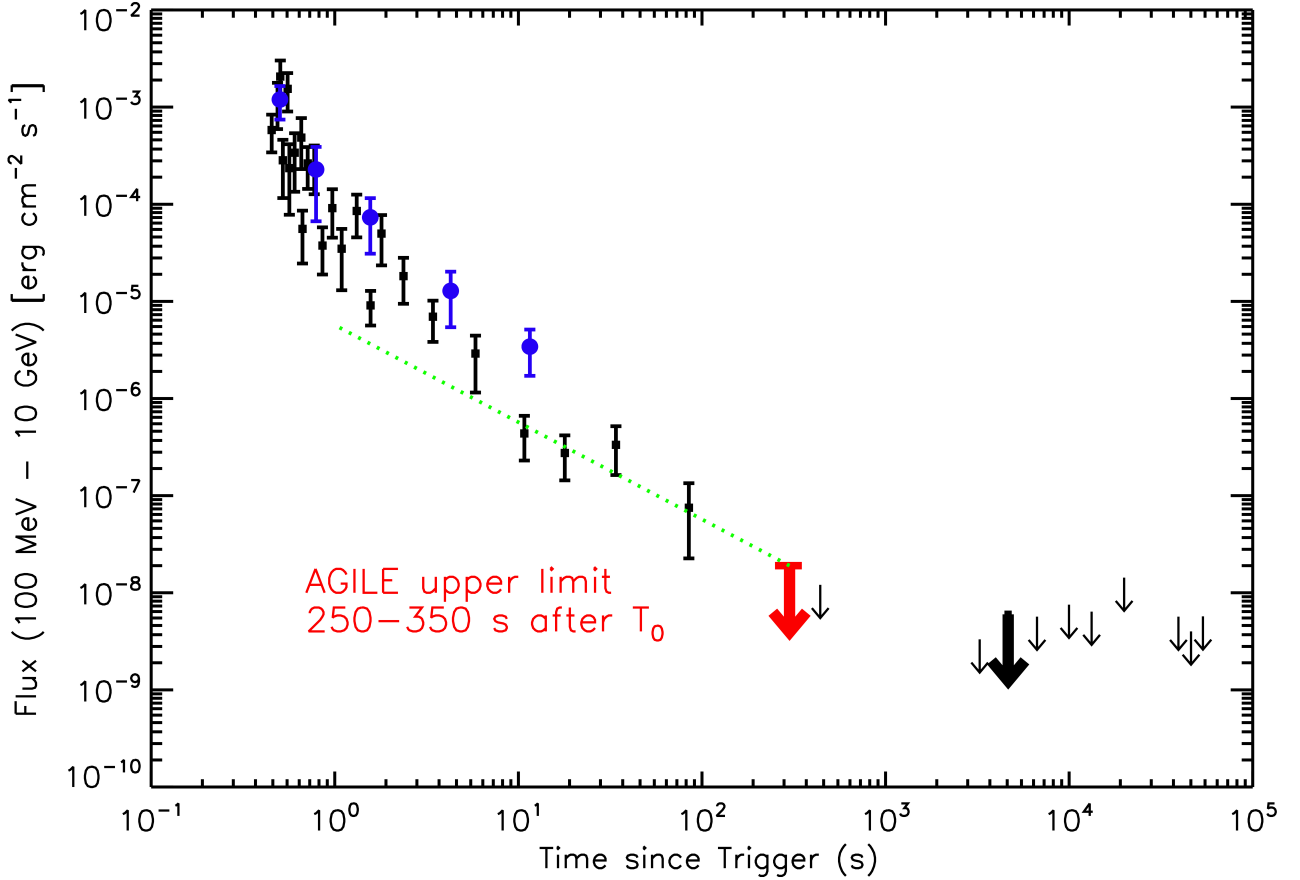

Fig. 6.— The AGILE (blue circles) and *Fermi*-LAT (black squares) gamma-ray lightcurves of the short GRB090510 (originally at $z = 0.9$) scaled in flux and time corrected as if it originated at the GW150914 luminosity distance of 400 Mpc ($z = 0.09$). The burst trigger time is assumed to be that of the GRB090510 precursor occurring 0.53 s before the main hard X-ray event (see Giuliani et al 2010, Ackermann et al. 2010). *Fermi*-LAT spectral data are from Fermi-LAT Collaboration (2016). The AGILE upper limit to gamma-ray emission above 100 MeV from the 65% of the GW150914 localization region during the time interval $\Delta T_{+1}$ is marked in red. The *Fermi*-LAT upper limit for GW150914 , obtained in the interval 4,442-4,867 s after the event, is marked in black color (*Fermi*-LAT Collaboration, 2016). The green dotted curve shows the estimated AGILE gamma-ray UL derived by extrapolating the UL near 300 s back to 1 s.

The probability of catching the "prompt" event in the GRID FoV is about 0.2 within the duration of the AGILE satellite spinning rotation. The overall probability for AGILE of having good exposure of the field of view during the prompt phase of a GW event is then $\sim 0.5 \times 0.2 = 0.1$. This is a relatively large probability compared to other satellites or ground instruments.

Further improvement of the efficiency and speed of the AGILE data processing is a task for the near future. The AGILE-GRID data management system allows to obtain results within $\sim$2 hours from the on-board data acquisition (Bulgarelli et al. 2014, Pittori et al. 2013). The data processing can be further optimized for a fast search of gamma-ray transients within a selected sky region communicated by an external alert. Super-AGILE, MCAL and AC data can be used in conjunction with GRID data to improve the search of transients. In addition, the reactivation of Super-AGILE data telemetry will make possible an event localization with 2 arcmin accuracy. The perspective for future follow-up gamma-ray observations of GW sources by AGILE is bright.

AGILE is an ASI space mission developed with programmatic support by INAF and INFN. We acknowledge partial support through the ASI grant no. I/028/12/0.